\documentclass[12pt,a4paper,twoside,english]{article}
\usepackage[T1]{fontenc}
\usepackage[latin1]{inputenc}
\usepackage{graphicx}
\usepackage{amssymb}

\makeatletter

\usepackage{babel}
\makeatother

\begin{document}

\title{Density oscillations in a model of water and other similar liquids}

\author{{\normalsize M. Apostol and E. Preoteasa }\\
{\normalsize Institute of Atomic Physics, }\\
{\normalsize Magurele-Bucharest MG-6, POBox MG-35, Romania }\\
{\normalsize email: apoma@theory.nipne.ro}}

\maketitle
\relax

\begin{abstract}
It is suggested that the dynamics of liquid water has a component
consisting of $O^{-2z}$ (oxygen) anions and $H^{+z}$ (hydrogen)
cations, where $z$ is a (small) reduced effective electron charge.
Such a model may apply to other similar liquids. The eigenmodes of
density oscillations are derived for such a two-species ionic plasma,
included the sound waves, and the dielectric function is calculated.
The plasmons may contribute to the elementary excitations in a model
introduced recently for the thermodynamics of liquids. It is shown
that the sound anomaly in water can be understood on the basis of
this model. The results are generalized to an asymmetric short-range
interaction between the ionic species as well as to a multi-component
plasma, and the structure factor is calculated.
\end{abstract}
\relax

\noindent \textbf{Introduction.} As simple as it may appear, water
is still a complex liquid involving various interactions as well as
kinematic and dynamic correlations. It is widely agreed that the water
molecule in liquid water preserves to some extent its integrity, especially
the directionality of the $sp^{3}$-oxygen orbitals, though it may
be affected substantially by hydrogen bonds.%
\footnote{L. Pauling, \emph{General Chemistry}, Dover, NY (1982); \emph{Water:
A Comprehensive Treatise}, ed. by F. Franks, Plenum, NY (1972).%
} As such, it is conceived that water has a molecular electric moment,
an intrinsic polarizability and hindered rotations (librations) which
may affect its orientational polarizability. We examine herein another
possible component of the dynamics of the liquid water, as resulting
from the dissociation of water molecule.

Water molecule $H_{2}O$ has two $H-O$ (hydrogen-oxygen) bonds which
make an angle of cca $109^{\circ}$ in accordance with the tetragonal
symmetry of the four hybridized $sp^{3}$-oxygen orbitals. The \char`\"{}spherical\char`\"{}
diameter of water molecule is approximately $2.75\textrm{\AA}$ and
the inter-molecular spacing in liquid water under normal conditions
is $a\sim3\textrm{\AA}$. This suggests that water molecule in liquid
water, while preserving the directionality of the oxygen electronic
orbitals, might be dissociated to a great extent. Dissociation models
which assume $OH^{-}-H^{+}$ or $OH^{-}-H_{3}O^{+}$ pairs are known
for water. This indicates a certain mobility of hydrogens (and oxygens).
We analyze herein the hypothesis that water may consist of $O^{-2z}$
anions of mass $M=16amu$ and density $n$ and $H^{+z}$ cations (protons)
of mass $m=1amu$ and density $2n$, where $z$ is a small reduced
effective electron charge (the atomic mass unit is $1amu\simeq1.7\times10^{-24}g.$).
We shall see that such a hypothesis adds another dimension to the
dynamics of water. Such a model may apply to other similar liquids. 

Due to their large mass the ions have a classical dynamics. Herein,
we limit ourselves to considering the ions motion in water under the
action of the Coulomb potentials $\varphi_{OO}=4z^{2}e^{2}/r$, $\varphi_{HH}=z^{2}e^{2}/r$
and $\varphi_{OH}=-2z^{2}e^{2}/r$, where $-e$ ($\simeq-4.8\times10^{-10}esu$)
is the electron charge and $r$ denotes the distance between the ions.
For stability, it is necessary also to introduce a short-range repulsive
(hard-core) potential $\chi$.%
\footnote{See also in this respect E. Teller, Revs. Mod. Phys. \textbf{34} 627
(1962); E. H. Lieb and B. Simon, Phys. Rev. Lett. \textbf{31} 681
(1973); Adv. Math. \textbf{23} 22 (1977); L. Spruch, Revs. Mod. Phys.
\textbf{63} 151 (1991). As it is well-known, a classical plasma with
Coulomb interaction only is unstable.%
} It is shown that in the limit $z\rightarrow0$ water may exhibit
an anomalous sound-like mode beside both the ordinary (hydrodynamic)
one and the non-equilibrium sound-like excitations governed by short-range
interactions. We compute the density oscillations for this model,
the dielectric function, the structure factor, and extend the model
to a multicomponent plasma, including an asymmetric short-range interaction
between ion species. 

\textbf{Plasmons in a jellium model.} Let us consider one species
of charged particles, with charge $-ze$, continuously distributed
with density $n$ in a neutralising rigid continuous background of
positive charge. This is the well-known jellium model.%
\footnote{See, for instance, D. Pines, \emph{Elementary Excitations in Solids},
Benjamin, NY (1963).%
} The Coulomb interaction reads \begin{equation}
U=\frac{1}{2}\int d\mathbf{r}d\mathbf{r}'\varphi(\mathbf{r}-\mathbf{r}')\delta n(\mathbf{r})\delta n(\mathbf{r}')\,\,\,,\label{1}\end{equation}
where $\delta n(\mathbf{r})$ denotes a small disturbance of density
(which preserves the global neutrality). We introduce the Fourier
representation \begin{equation}
\delta n(\mathbf{r})=\frac{1}{\sqrt{N}}\sum_{\mathbf{q}}\delta n(\mathbf{q})e^{i\mathbf{qr}}\,\,\,,\,\,\,\delta n(\mathbf{q})=\frac{n}{\sqrt{N}}\int d\mathbf{r}\delta n(\mathbf{r})e^{-i\mathbf{qr}}\,\,\,,\label{2}\end{equation}
 where $N=nV$ is the total number of particles in volume $V$. Similarly,
\begin{equation}
\varphi(\mathbf{r})=\frac{1}{V}\sum_{\mathbf{q}}\varphi(\mathbf{q})e^{i\mathbf{qr}}\,\,\,,\,\,\,\varphi(\mathbf{q})=\int d\mathbf{r}\varphi(\mathbf{r})e^{-i\mathbf{qr}}\,\,\,,\label{3}\end{equation}
where $\varphi(\mathbf{q})=4\pi z^{2}e^{2}/q^{2}$ is the Fourier
transform of the Coulomb potential (interaction). The Coulomb interaction
given by (\ref{1}) becomes \begin{equation}
U=\frac{1}{2n}\sum_{\mathbf{q}}\varphi(q)\delta n(\mathbf{q})\delta n(-\mathbf{q})\,\,\,\label{4}\end{equation}
(where the $q=0$-term is excluded by the positive background). 

The small variations $\delta n(\mathbf{r})$ in density can be represented
as $\delta n=-ndiv\mathbf{u}$, where $\mathbf{u}$ is a displacement
vector.%
\footnote{M. Apostol, \emph{Electron Liquid}, apoma, MB (2000).%
} We emphasize that such a representation holds for $\mathbf{qu}(\mathbf{r})\ll1$.
It follows $\delta n(\mathbf{q})=-in\mathbf{q}\mathbf{u(}\mathbf{q})$,
and one can see that the Coulomb interaction involves only longitudinal
components of the displacement vector $\mathbf{u}(\mathbf{q})$ along
the wavevector $\mathbf{q}$. Therefore, we may write $\mathbf{u}(\mathbf{q})=(\mathbf{q}/q)u(\mathbf{q})$,
with $\delta n^{*}(-\mathbf{q})=\delta n(\mathbf{q})$, $\mathbf{\mathbf{u^{*}(-}q)\mathbf{=}u(}\mathbf{q})$
and $u^{*}(-\mathbf{q})=-u(\mathbf{q})$. The Coulomb interaction
(\ref{4}) becomes \begin{equation}
U=-\frac{n}{2}\sum_{\mathbf{q}}q^{2}\varphi(q)u(\mathbf{q})u(-\mathbf{q})\,\,.\label{5}\end{equation}

The kinetic energy associated with the coordinates $u(\mathbf{q})$
is given by \begin{equation}
T=\frac{1}{2}\int d\mathbf{r}nm\dot{\mathbf{u}}^{2}=-\frac{1}{2}m\sum_{\mathbf{q}}\dot{u}(\mathbf{q})\dot{u}(-\mathbf{q})\,\,\,,\label{6}\end{equation}
where $m$ denotes the particle mass. The equations of motion obtained
from the Lagrange function $L=T-U$ are \begin{equation}
m\ddot{u}(\mathbf{q})+nq^{2}\varphi(q)u(\mathbf{q})=0\,\,\,,\label{7}\end{equation}
 which leads to the well-known plasma oscillations with frequency
given by $\omega_{p}^{2}=4\pi nz^{2}e^{2}/m$.

\textbf{Plasma oscillations with two species of ions.} We apply the
above model to the two species of ions $O^{-2z}$ and $H^{+z}$. The
change in density is associated with a displacement vector $\mathbf{v}$
in the former and a displacement vector $\mathbf{u}$ in the latter.
First we note that the Fourier transforms of the Coulomb potentials
are given by $\varphi_{OO}=4\varphi(q)$, $\varphi_{HH}=\varphi(q)$
and $\varphi_{OH}=-2\varphi(q)$, where $\varphi(q)=4\pi z^{2}e^{2}/q^{2}$.
Therefore, the interactions can be written as \begin{equation}
\begin{array}{c}
U_{OO}=-\frac{n}{2}\sum_{\mathbf{q}}q^{2}\left[4\varphi(q)+\chi(q)\right]v(\mathbf{q})v(-\mathbf{q})\,\,\,,\\
\\U_{HH}=-2n\sum_{\mathbf{q}}q^{2}\left[\varphi(q)+\chi(q)\right]u(\mathbf{q})u(-\mathbf{q})\,\,\,,\\
\\U_{OH}=n\sum_{\mathbf{q}}q^{2}[2\varphi(q)-\chi]u(\mathbf{q})v(-\mathbf{q})\,\,\,,\end{array}\label{8}\end{equation}
 where $n=N/V$ is the density of water molecules and the Fourier
transform $\chi$ of a hard-core potential has been introduced (the
same for both species). The kinetic energy is given by \begin{equation}
T=-\frac{1}{2}M\sum_{\mathbf{q}}\dot{v}(\mathbf{q})\dot{v}(-\mathbf{q})-m\sum_{\mathbf{q}}\dot{u}(\mathbf{q})\dot{u}(-\mathbf{q})\,\,\,,.\label{9}\end{equation}
and the equations of motion read \begin{equation}
\begin{array}{c}
m\ddot{u}+2nq^{2}(\varphi+\chi)u-nq^{2}(2\varphi-\chi)v=0\,\,\,\\
\\M\ddot{v}+nq^{2}(4\varphi+\chi)v-2nq^{2}(2\varphi-\chi)u=0\,\,\,,\end{array}\label{10}\end{equation}
where we have dropped out the argument $\mathbf{q}$. 

The solutions of these equations can be obtained straightforwardly.
In the long wavelength limit $\mathbf{q}\rightarrow0$ there are two
branches of eigenfrequencies, one given by \begin{equation}
\omega_{p}^{2}=\frac{16\pi nz^{2}e^{2}}{\mu}\,\,\,\label{11}\end{equation}
 corresponding to plasma oscillations and another given by \begin{equation}
\omega_{s}^{2}=\frac{9n\chi}{M+2m}q^{2}=v_{s}^{2}q^{2}\,\,\,\label{12}\end{equation}
 corresponding to sound-like waves propagating with velocity $v_{s}$
given by (\ref{12}). $\mu=2mM/(2m+M)$ is the reduced mass. The plasma
oscillations are associated with antiphase oscillations of the relative
coordinate ($2mu+Mv=0$), while the sound waves are associated with
in-phase oscillations of the center-of-mass coordinate ($u-v=0$).

\textbf{Polarization. }An external electric field arising from a potential
$\phi(\mathbf{r})$ gives an additional energy \begin{equation}
U_{i}=q_{i}\int d\mathbf{r}\phi(\mathbf{r})\delta n_{i}(\mathbf{r})=-i(n_{i}q_{i}/n)\sum_{\mathbf{q}}q\phi(\mathbf{q})u_{i}(-\mathbf{q})\,\,\,,\label{13}\end{equation}
for a species of ions labelled by $i$, with electric charge $q_{i}$
and density $n_{i}$. We apply this formula to the two-species ionic
plasma, and get \begin{equation}
U_{H}=-2ize\sum_{\mathbf{q}}q\phi(\mathbf{q})u(-\mathbf{q})\,\,,\,\, U_{O}=2ize\sum_{\mathbf{q}}q\phi(\mathbf{q})v(-\mathbf{q})\,\,\label{14}\end{equation}

Adding these two terms to the lagrangian, the equations of motion
given by (\ref{10}) become \begin{equation}
\begin{array}{c}
m\ddot{u}+2nq^{2}(\varphi+\chi)u-nq^{2}(2\varphi-\chi)v=-izeq\phi\,\,\,\\
\\M\ddot{v}+nq^{2}(4\varphi+\chi)v-2nq^{2}(2\varphi-\chi)u=2izeq\phi\,\,\,,\end{array}\label{15}\end{equation}
where we have dropped out the argument $\mathbf{q}$. This is a system
of coupled harmonic oscillators under the action of an external force.
In the limit of long wavelengths its solutions are given by \begin{equation}
\begin{array}{c}
u=\frac{izeq}{m}\phi\frac{\omega^{2}-\frac{2m}{3\mu}\omega_{s}^{2}}{(\omega^{2}-\omega_{p}^{2})(\omega^{2}-\omega_{s}^{2})}\,\,\,,\\
\\v=-\frac{2izeq}{M}\phi\frac{\omega^{2}-\frac{2M}{3\mu}\omega_{s}^{2}}{(\omega^{2}-\omega_{p}^{2})(\omega^{2}-\omega_{s}^{2})}\,\,.\end{array}\label{16}\end{equation}
On the other hand, equation $n_{i}div\mathbf{u}_{i}=-\delta n_{i}$
is in fact the Maxwell equation $div\mathbf{E}_{i}=4\pi q_{i}\delta n_{i}$,
where the electric field is given by $\mathbf{E}_{i}=-4\pi nq_{i}\mathbf{u}_{i}$.
We have therefore the internal electric fields $E_{u}=-8\pi nzeu$
and $E_{v}=8\pi nzev$. The polarization $P=-(E_{u}+E_{v})/4\pi$
is given by \begin{equation}
P(\mathbf{q})=2nze\left[u(\mathbf{q})-v(\mathbf{q})\right]=\frac{iq}{4\pi}\phi(\mathbf{q})\frac{\omega_{p}^{2}}{\omega^{2}-\omega_{p}^{2}}\,\,.\label{17}\end{equation}
 The external field is related to the external potential through $D(\mathbf{q})=-iq\phi(\mathbf{q})$
and the dielectric function $\varepsilon$ is given by $D=\varepsilon E=\varepsilon(D+E_{int})$,
where $E_{int}=E_{u}+E_{v}$ is the internal field. We get the dielectric
function%
\footnote{We disregard here the intrinsic and orientational polarizabilities.%
} \begin{equation}
\varepsilon=1-\omega_{p}^{2}/\omega^{2}\,\,\,,\label{18}\end{equation}
as expected. As it is well-known, its zero gives the longitudinal
mode of plasma oscillations. 

The $\omega_{p}$ in the nominator of equation (\ref{18}) defines
also the plasma edge: for frequencies lower than $\omega_{p}$ the
electromagnetic waves are absorbed (the refractive index is given
by $n^{2}=\varepsilon$). It is well-known that water exhibits indeed
a strong absorption in the gigahertz-terrahertz region.%
\footnote{See, for instance, K. H. Tsai and T.-M. Wu, Chem. Phys. Lett. \textbf{417}
390 (2005); A. Padro and J. Marti, J. Chem. Phys. \textbf{118} 452
(2003); K. N. Woods and H. Wiedemann, Chem. Phys. Lett. \textbf{393}
159 (2004).%
} On the other hand, neutron scattering on heavy water,%
\footnote{F. J. Bermejo, M. Alvarez, S. M. Bennington and R. Vallauri, Phys.
Rev. \textbf{E51} 2250 (1995); C. Petrillo, F. Sacchetti, B. Dorner
and J.-B. Suck, Phys. Rev. \textbf{E62} 3611 (2000).%
} as well as inelastic $X$-ray scattering,%
\footnote{F. Sette, G. Ruocco, M. Krisch, C. Masciovecchio, R. Verbeni and U.
Bergmann, Phys. Rev. Lett. \textbf{77} 83 (1996).%
} revealed the existence of a dispersionless mode $\simeq4-5meV$ ($\simeq10^{13}s^{-1}$)
in the structure factor, which may be taken tentatively as the $\omega_{p}$-plasmonic
mode given by equation (\ref{11}). Making use of this equation we
get $\omega_{p}\simeq3\times10^{14}zs^{-1}$($n=1/a^{3},$ $a=3\textrm{\AA}$),
so we may estimate the reduced effective charge $z\simeq3\times10^{-2}$. 

\textbf{Dielectric function.} The dielectric function given by equation
(\ref{18}) has a singularity for $\omega=0$, as arising from the
exact cancellation in the static limit of the external field by the
internal field. It is plausible to assume that residual polarization
fields are still present in this static limit, like, for instance,
the intrinsic polarizability. In this case, equation (\ref{18}) is
modified, and the dielectric function is of the type \begin{equation}
\varepsilon=\frac{\omega^{2}-\omega_{p}^{2}}{\omega^{2}+\omega_{0}^{2}}\,\,\,,\label{19}\end{equation}
where $\omega_{0}$ is a plasma frequency associated with the intrinsic,
molecular polarizability.%
\footnote{A static field $D$ produces an electric dipole $p=q_{e}x$, where
$q_{e}$ is the electric charge and $x$ is a small displacement subjected
to the equation of motion $m_{e}\ddot{x}+m_{e}\omega_{p}^{2}x=q_{e}D$,
where $m_{e}$ is the mass of the electronic cloud. According to the
plasma model suggested here, we assume that the electronic cloud in
the $H-O$ bonds have the same eigenfrequency $\omega_{p}$ as the
$H-O$ ensemble. In the static limit $x=q_{e}D/m_{e}\omega_{p}^{2}$
(polarizability $\alpha=q_{e}^{2}/m_{e}\omega_{p}^{2}$ in $p=\alpha D$),
and we get a polarization $P=p/a_{0}^{3}=q_{e}^{2}D/m_{e}a_{0}^{3}\omega_{p}^{2}$,
where $a_{0}$ is of the order of the atomic size. We get an internal
field $E_{int}=-4\pi P=-\left(4\pi q_{e}^{2}/m_{e}a_{0}^{3}\right)D/\omega_{p}^{2}=-\left(\omega_{0}^{2}/\omega_{p}^{2}\right)D$,
where $\omega_{0}$ is a frequency of the order of atomic frequencies.
Consequently, the dielectric function $\varepsilon$ in equation $D=\varepsilon E=\varepsilon(D+E_{int})$
is given by $\varepsilon\simeq-\omega_{p}^{2}/\omega_{0}^{2}$ ($\omega_{p}^{2}/\omega_{0}^{2}\ll1$),
which is precisely the static dielectric function given by equation
(\ref{19}). %
} As such, it is a very high frequency, and equation (\ref{19}) gives
a small, negative contribution to the dielectric function in the static
limit ($\omega\rightarrow0$). 

The dielectric properties of water are still a matter of debate. It
is agreed that the permitivity dispersion of water is described to
some extent by a Debye model of the form $\varepsilon=a+b/(1-i\omega\tau)$,
where $a$ and $b$ are semi-empirical parameters and $\tau\sim\eta a^{3}/T$
is a relaxation time; $\eta$ denotes the viscosity and $T$ is the
temperature.%
\footnote{See, for instance, H. Frohlich, \emph{Theory of Dielectrics}, Oxford
(1958); P. Debye, \emph{Polar Molecules}, Dover, NY (1945).%
} This Debye model assumes mainly an orientational polarizability of
electric dipoles, which, due to the preservation of the directional
character of the $O-H$ bonds, is compatible with the plasma model
suggested here for water. Therefore, the contribution given by equation
(\ref{19}) should be added to the above Debye formula for the dielectric
function, which becomes \begin{equation}
\varepsilon=a+\frac{b}{1-i\omega\tau}+\frac{\omega^{2}-\omega_{p}^{2}}{\omega^{2}+\omega_{0}^{2}}\,\,.\label{20}\end{equation}
Parameters $a$ and $b$ in equation (\ref{20}) are related to the
static permitivity $\varepsilon_{0}$ and high-frequency permitivity
$\varepsilon_{\infty}$ through \begin{equation}
\varepsilon_{0}=a+b-\omega_{p}^{2}/\omega_{0}^{2}\,\,,\,\,\varepsilon_{\infty}=a+1\,\,.\label{21}\end{equation}
We may neglect $\omega_{p}^{2}/\omega_{0}^{2}$ here because it is
too small, and we may also take $\varepsilon_{\infty}=1$($a=0$).
The static permitivity $\varepsilon_{0}=b$ is given mainly by the
electric dipoles. Let $\mathbf{p}$ be such an electric dipole. Its
energy in an electric field $\mathbf{D}$ is $-pD\cos\theta$, where
$\theta$ is the angle between $\mathbf{p}$ and $\mathbf{D}$. The
thermal distribution of such dipoles is $dw\sim\exp(-pD\cos\theta/T)d(\cos\theta),.$where
$T$ denotes the temperature. We get easily the thermal average $\left\langle \cos\theta\right\rangle =-L(pD/T)$,
where $L(x)=\coth x-1/x$ is the well-known Langevin's function. 

We take $p=2ez_{e}(a/2)=ez_{e}a$, where $a\sim3\textrm{\AA}$ and
$z_{e}$ is a delocalized reduced charge associated with the $H-O$
dipole. We estimate the argument $pD/T$ of the Langevin's function.
At room temperature, we find $pD/T\simeq3\times10^{-4}Dz_{e}$. For
$pD/T=1$ this corresponds to an external field $D=\frac{1}{3z_{e}}\times10^{4}esu$,
or $D=10^{8}/z_{e}V/m$.%
\footnote{$1esu=3\times10^{4}V/m$, J. D. Jackson, \emph{Classical Electrodynamics},
Wiley, NJ (1999).%
} This is an extremely high field, so we are justified to take $pD/T\ll1$,
and $L(pD/T)\simeq pD/3T$. We get therefore a polarization $P=-np\left\langle \cos\theta\right\rangle =np^{2}D/3T$,
an internal field $E_{int}=-4\pi P=-4\pi np^{2}D/3T$, and a permitivity
\begin{equation}
\varepsilon_{0}=b=\frac{1}{1-4\pi np^{2}/3T}\,\,\,\label{22}\end{equation}
from $D=\varepsilon E=\varepsilon(D+E_{int})$. This is the well-known
Kirkwood formula.%
\footnote{See, for instance, H. Frohlich, \emph{loc cit}.%
} For the empirical value $\varepsilon_{0}=80$, we get (at room temperature)
a reduced charge $z_{e}\simeq10^{-2}$. This is in good agreement
with the $H^{+z}-O^{-2z}$ plasma charge $z$ estimated above. 

\textbf{Cohesion and thermodynamics.} Recently, a model of liquid
has been introduced%
\footnote{M. Apostol, J.Theor. Phys. \textbf{125} 163 (2006).%
} based on an excitation spectrum (per particle) of the form $\varepsilon_{n}=-\varepsilon_{0}+\varepsilon_{1}(n+1/2)$,
where $\varepsilon_{0}$ is a cohesion energy and $\varepsilon_{1}$
is the quanta of energy of a harmonic oscillator with one degree of
freedom; $n$ represents here the quantum number. The model includes
also the kinematic correlations (spatial restrictions) of the movement
of the liquid molecules. This model leads to a consistent thermodynamics
for liquids, arising from a statistics which is equivalent with the
statistics of bosons in two dimensions. 

For water, the cohesion energy per particle $\varepsilon_{0}$ can
be estimated from the vaporization heat ($\simeq40kJ/mol$). It gives
$\varepsilon_{0}\sim10^{3}K$. On the other hand, it was shown in
a previous paper%
\footnote{M. Apostol, Mod. Phys. Let. \textbf{B21} 893 (2007); see also M. Apostol,
J. Theor. Phys. \textbf{123} 155 (2006).%
} that the transition temperature beween a gas and a liquid of identical
particles is approximately given by \begin{equation}
T_{t}=\frac{4}{3}\frac{\varepsilon_{0}}{\ln(\varepsilon_{0}/T_{0})}\,\,\,,\label{23}\end{equation}
 where $T_{0}=\hbar^{2}n^{2/3}/m$ is a gas characteristic temperature.
We can apply this formula to water disssociation, taking $n$ as the
density of hydrogen atoms, $m$ as the mass of two hydrogen atoms
and $T_{t}=383K$ (at normal pressure; $\varepsilon_{0}$ depends
on the inter-particle spacing). We may neglect the oxygen, as it is
too heavy in comparison with the hydrogen atoms. We get $T_{0}\simeq2K$
and the above formula gives $\varepsilon_{0}\simeq2000K\simeq200meV$
for the cohesion energy of water per molecule, which is consistent
with the above estimate ($1eV\simeq11.6\times10^{3}K$; $n\simeq1/a^{3}$
with $a=3\textrm{\AA}$ and $\hbar\simeq10^{-27}erg\cdot s$; Bohr
radius $a_{H}=\hbar^{2}/m_{e}e^{2}\simeq0.53\textrm{\AA}$, $e^{2}/a_{H}\simeq27.2eV$,
where $m_{e}$ is the electron mass).%
\footnote{It is worth noting that the mechanism of vaporization assumed here
implies the dissociation of the water molecule.%
} 

The plasma oscillations obtained above can be quantized and the energy
levels of the plasma read \begin{equation}
E_{n}=\sum_{\mathbf{q}}\hbar\omega_{p}(n+1/2)=\frac{V}{(2\pi)^{3}}\frac{4\pi}{3}q_{c}^{3}\cdot\hbar\omega_{p}(n+1/2)\,\,\,,\label{24}\end{equation}
 where $q_{c}$ is a cutoff wavevector. The prefactor in equation
(\ref{24}) is $Vq_{c}^{3}/6\pi^{2}\simeq N(aq_{c}/4)^{3}$, so the
energy levels given above can be written as \begin{equation}
E_{n}=N\varepsilon_{1}(n+1/2)\,\,\,,\label{25}\end{equation}
 where $\varepsilon_{1}=(aq_{c}/4)\hbar\omega_{p}$. These energy
levels correspond to a harmonic oscillator with one degree of freeedom.
It follows that the present description of water as a two-species
of highly dissociated ionic plasma provides a further support for
the liquid model mentioned above. If we take $q_{c}\simeq1/a$ the
energy quanta $\varepsilon_{1}=(aq_{c}/4)^{3}\hbar\omega_{p}=\simeq3zmeV$
represents the $\varepsilon_{1}$ parameter in the spectrum of the
liquid. (The plasma frequency given by equation (\ref{11}) is $\omega_{p}\simeq200zmeV$).

\textbf{Debye screening and the correlation energy.} As it is well-known
the plasma excitations described above represent collective oscillations
of the density in the long wavelength limit. At the same time they
induce correlations in the ionic movements. For a classical plasma
these correlations are associated with a screening length given by
the Debye-Huckel theory as %
\footnote{See, for instance, L. Landau and E. Lifshitz, \emph{Course of Theoretical
Physics}, vol. 5, \emph{Statistical Physics}, Elsevier (1980).%
} \begin{equation}
\kappa^{-1}=\left(T/24\pi nz^{2}e^{2}\right)^{1/2}\,\,\,\label{26}\end{equation}
for our case ($\kappa^{-1}=\left(T/4\pi e^{2}\sum_{i}n_{i}z_{i}^{2}\right)^{-1}$where
$i$ labels the ionic species with density $n_{i}$ and charge $ez_{i}$).
The formula is valid for the Coulomb energy $z^{2}e^{2}/a$ much lower
than the temperature $T$. In the present case we have $z^{2}e^{2}/a\simeq45K$
(for $z\simeq3\times10^{-2}$), which shows that the above condition
is fulfilled. From (\ref{26}) we get $\kappa^{-1}\sim1\textrm{\AA}$
(at room temperature), in agreement with the present molecular-dissociation
model. The correlation energy per particle is given by \begin{equation}
\varepsilon_{corr}=-\frac{e^{2}}{a}\sqrt{\frac{\pi e^{2}}{Ta}}(6z^{2})^{3/2}\,\,\,\label{27}\end{equation}
($\varepsilon_{corr}=-(e^{2}/a)\sqrt{\pi e^{2}/Ta}(\sum_{i}n_{i}z_{i}^{2})^{3/2}$).
The estimation of this energy gives $\varepsilon_{corr}\sim10^{2}K$
(at room temperature). It contributes to the cohesion energy. 

\textbf{Sound anomaly.} The sound-like branch $\omega_{2}\simeq\omega_{s}=v_{s}q$,
where $v_{s}=\sqrt{9n\chi/(M+2m)}$ according to equation (\ref{12}),
is distinct from the ordinary hydrodynamic sound whose velocity is
given by the well-known formula $v_{0}=1/\sqrt{\kappa nm}$ for a
one-component fluid, where $\kappa$ is the adiabatic compressibility.
For the present two-component fluid ($H^{+z}-O^{-2z}$ plasma), the
velocity of the ordinary sound is given by $v_{0}=1/\sqrt{\kappa n(M+2m)}$.
The former represents a non-equilibrium elementary excitation, whose
velocity $v_{s}$ does not depend on temperature, while the latter
proceeds by thermodynamic, equilibrium, adiabatic processes, and its
velocity $v_{0}$ depends on temperature thorugh the adiabatic conpresibility
$\kappa$. In order to distinguish them from the hydrodynamic sound
we propose to call the sound-like excitations derived here density
\char`\"{}kinetic\char`\"{} modes or \char`\"{}densitons\char`\"{}.
The distinction between the two sounds is made by a threshold wavevector
$q_{t}$ in the following manner. Suppose that there is a finite lifetime
$\tau$ for the sound-like excitations $\omega_{s}$ propagating with
a velocity $v_{s}$ and a corresponding meanfree path $\Lambda=v_{s}\tau$.
If the sound-like wavelength $\lambda$ is much longer than the meanfree
path, $\lambda\gg\Lambda$, then we are in the collision-like regime
($\omega_{s}\tau\ll1$), and the collisions may restore the thermodynamic
equilibrium. In this case the hydrodynamic sound propagates, and the
sound-like excitations do not. This condition defines the threshold
wavevector $q_{t}=1/v_{s}\tau$. In the opposite case, $q\gg q_{t}$
(collision-less regime), it is the sound-like excitations that propagate,
and not the hydrodynamic sound. The finite lifetime $\tau$ originates
in the residual interactions between the collective modes and the
underlying motion of the individual particles. It is easy to estimate
this residual interaction.%
\footnote{M. Apostol, \emph{Electron Liquid}, apoma, MB (2000). %
} It is given by $\sqrt{\varepsilon T}$, where $\varepsilon$ is the
mean energy per particle corresponding to the motion of the individual
particles. We get therefore $\tau\simeq\hbar/\sqrt{\varepsilon T}$
and the threshold wavevector $q_{t}=\sqrt{\varepsilon T}/\hbar v_{s}$.
It is difficult to have a reliable estimation of the mean energy $\varepsilon$;
for a resonable value $\varepsilon=10meV$ we get $q_{t}\simeq0.1\textrm{\AA}^{-1}$
at room temperature for $v=3000m/s$, which is in good agreement with
experimental data.

Indeed, the phenomenon of two-sound anomaly in water is well-documented.%
\footnote{See, for instance, J. Teixeira, M. C. Bellissent-Funel, S. H. Chen
and B. B Dorner, Phys. Rev. Lett. \textbf{54} 2681 (1985); S. C. Santucci,
D. Fioretto, L. Comez, A. Gessini and C. Maschiovecchio, Phys. Rev.
Lett. \textbf{97} 225701 (2006) and references therein.%
} Neutron, $X$-ray, Brillouin or ultraviolet light scattering on water
revealed the existence of a hydrodynamic sound propagating with velocity
$v_{0}\simeq1500m/s$ for smaller wavevectors and an additional sound
propagating with velocity $\simeq3000m/s$ for larger wavevectors.
In addition, though both sound velocities do exhibit an isotopic effect,
their ratio does not. According to the above discussion, we assign
this additional, faster sound to the sound-like excitations derived
here. We can see that both $v_{0}$ and $v_{s}$ given above exhibit
a weak isotopic effect, while their ratio $v_{s}/v_{0}=3n\sqrt{\kappa\chi}$
does not. From $v_{s}=\sqrt{9n\chi/(M+2m)}=3000m/s$ we get the short-range
interaction $\chi\simeq7eV\cdot\textrm{\AA}^{3}$. Similar results
are obtained for other forms of dissociation of the water molecule,
like $OH^{-}-H^{+}$ or $OH^{-}-H_{3}O^{+}$, so the $H^{+z}-O^{-2z}$
plasma model employed here can be viewed as an average, effective
model for various plasma components that may exist in water.

\textbf{Another possible anomalous sound.} It is worth calculating
the spectrum given by equations of motion (\ref{10}) without neglecting
higher-order contributions in $q^{2}$. The result of this calculation
is given by \begin{equation}
\omega_{1,2}^{2}=\frac{1}{2}\omega_{p}^{2}\left[1+Ax^{2}\pm\sqrt{1+2Bx^{2}+A^{2}x^{4}}\right]\,\,\,,\label{28}\end{equation}
 where \begin{equation}
A=\frac{1}{9\alpha}(2+5\alpha+2\alpha^{2})\,\,,\,\, B=\frac{1}{9\alpha}(2-13\alpha+2\alpha^{2})\,\,,\,\,\alpha=m/M\,\,\,\label{29}\end{equation}

and $x=v_{s}q/\omega_{p}$. It is shown in Fig. 1.%
\begin{figure}
\noindent \begin{centering}
\includegraphics[clip,scale=0.4]{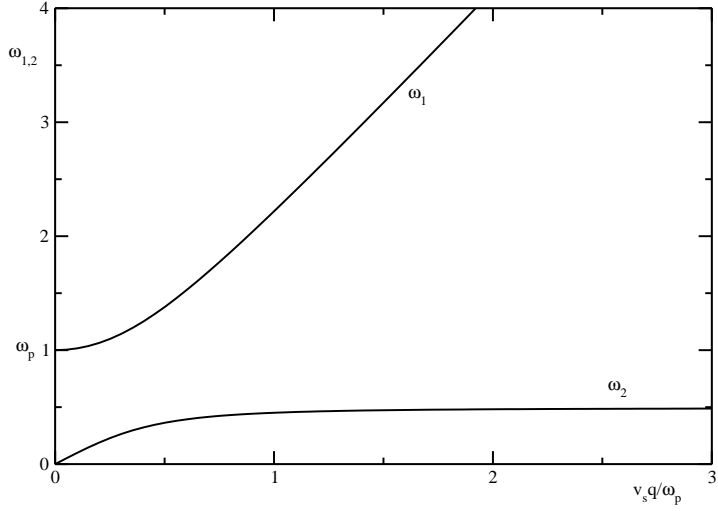}
\par\end{centering}

\caption{The spectrum of the density oscillations given by equation (\ref{28})
for the $H^{+z}-O^{-2z}$ plasma with the same short-range interaction
between ionic species.}

\end{figure}

Frequency $\omega_{2}$ in equation (\ref{28}) represents the sound-like
branch, which goes like $\omega_{2}\simeq\omega_{s}=v_{s}q$ in the
long wavelength limit and approaches the horizontal asymptote $\omega_{2}=\omega_{p}/\sqrt{A}\simeq\omega_{p}\sqrt{m/2M}$
for shorter wavelengths. Frequency $\omega_{1}$ in equation (\ref{28})
represents the plasmonic branch ($\omega_{1}\simeq\omega_{p}$ for
$q\rightarrow0$). In the long wavelength limit it goes like \begin{equation}
\omega_{1}\simeq\omega_{p}+\frac{(M-m)^{2}}{9mM}v_{s}^{2}q^{2}/\omega_{p}\,\,\,,\,\,\, q\rightarrow0\,\,.\label{30}\end{equation}
 Due to the large disparity between the two masses $m$ and $M$ we
can see that the plasma frequency has an abrupt increase toward the
short-wavelength oblique asymptote given by \begin{equation}
\omega_{a}\simeq\sqrt{A}v_{s}q\simeq\sqrt{2M/9m+5/9}v_{s}q\,\,.\label{31}\end{equation}

For small values of $\omega_{p}$ (vanishing Coulomb coupling, $z\rightarrow0$)
this asymptotic frequency may look like an anomalous sound propagating
with velocity \begin{equation}
v_{a}\simeq\sqrt{2M/9m+5/9}v_{s}\,\,.\label{32}\end{equation}
For water, we get $v_{a}\simeq2v_{s}$ from this formula. However,
the ratios $v_{a}/v_{s}$ or $v_{a}/v_{0}$ exhibit a rather strong
isotopic effect, which is not supported by experimental data. 

\textbf{Multi-component plasma.} The model presented herein might
be generalized to a multi-component plasma consisting of several ionic
species labelled by $i$, each with number $N_{i}$ of particles,
density $n_{i}$, charge $z_{i}e$ and mass $m_{i}$, such that $\sum_{i}z_{i}n_{i}=0$. 

The lagrangian of the density oscillations is given by\begin{equation}
\begin{array}{c}
L=-\frac{1}{2n}\sum_{i\mathbf{q}}m_{i}n_{i}\dot{u}_{i}(\mathbf{q})\dot{u}_{i}(-\mathbf{q})+\frac{1}{2n}\sum_{ij\mathbf{q}}n_{i}n_{j}q^{2}\left[\varphi_{ij}(q)+\chi(q)\right]u_{i}(\mathbf{q})u_{j}(-\mathbf{q})+\\
\\+i\frac{e}{n}\sum_{i\mathbf{q}}n_{i}z_{i}q\phi(\mathbf{q})u_{i}(-\mathbf{q})\,\,\,,\end{array}\label{33}\end{equation}
where $\varphi_{ij}(q)=4\pi z_{i}z_{j}e^{2}/q^{2}$. The equations
of motion are given by \begin{equation}
m_{i}\ddot{u}_{i}+4\pi e^{2}z_{i}\sum_{j}z_{j}n_{j}u_{j}+q^{2}\chi\sum_{j}n_{j}u_{j}=-iqez_{i}\phi\,\,.\label{34}\end{equation}
 Making use of the notations \begin{equation}
S_{1}=\sum_{i}z_{i}^{2}n_{i}/m_{i}\,\,,\,\, S_{2}=\sum_{i}n_{i}/m_{i}\,\,,\,\, S_{3}=\sum_{i}z_{i}n_{i}/m_{i}\,\,\,,\label{35}\end{equation}
the eigenfrequencies $\omega_{1,2}$ of the system of equations (\ref{34})
in the long wavelength limit are given by \begin{equation}
\omega_{1}^{2}\simeq\omega_{p}^{2}=4\pi e^{2}S_{1}=\sum_{i}\frac{4\pi e^{2}z_{i}^{2}n_{i}}{m_{i}}\,\,\,,\label{36}\end{equation}
 which represents the plasma branch of the spectrum, and \begin{equation}
\omega_{2}^{2}\simeq\omega_{s}^{2}=\left(S_{2}-S_{3}^{2}/S_{1}\right)\chi q^{2}=v_{s}^{2}q^{2}\,\,\,,\label{37}\end{equation}
which represents the sound-like excitations.%
\footnote{The sound velocity given by (\ref{37}) is always a real quantity,
as a consequence of the Schwarz-Cauchy inequality.%
} The plasma branch of the spectrum has an oblique asimptote given
by $\omega_{1}\simeq\omega_{a}=\sqrt{\chi S_{2}}q$, which may be
taken as an anomalous sound propagating with velocity $v_{a}=\sqrt{\chi S_{2}}$
for small values of $\omega_{p}$. The ratio of the two sound velocities
is given by \begin{equation}
v_{a}/v_{s}=\frac{1}{\sqrt{1-S_{3}^{2}/S_{1}S_{2}}}\,\,\,,\label{38}\end{equation}
which is always higher than unity. The sound branch of the spectrum
has an horizontal asymptote given by $\omega_{2}\simeq\sqrt{1-S_{3}^{2}/S_{1}S_{2}}\omega_{p}$.
For the $H^{+z}-O^{-2z}$ plasma we can check from (\ref{38}) that
$v_{a}/v_{s}\simeq(2M/9m+5/9)^{1/2}\simeq2$, and $\omega_{2}\simeq3\sqrt{m/2M}\omega_{p}$,
as obtained above. As we have discussed above this ratio exhibits
a rather strong isotopic effect, which is not in accord with experimental
data. We assign therefore the additional sound to sound-like excitations
propagating with velocity $v_{s}$ given by equation (\ref{37}).
The ordinary, hydrodynamic sound in a multi-component mixture has
the velocity $v_{0}=1/\sqrt{\kappa\sum_{i}n_{i}m_{i}}$. It can be
shown that $v_{s}^{2}/v_{0}^{2}\geq n^{2}\kappa\chi$ for a neutral
multi-component mixture. 

The internal field is given by \begin{equation}
E_{int}=-4\pi e\sum_{i}z_{i}n_{i}u_{i}\,\,\,;\label{39}\end{equation}
 we get easily from equations (\ref{34}) \begin{equation}
E_{int}=-iq\phi\frac{\omega_{p}^{2}}{\omega^{2}-\omega_{p}^{2}}\,\,\,\label{40}\end{equation}
and the dielectric function $\varepsilon=1-\omega_{p}^{2}/\omega^{2}$,
as expected. 

\textbf{Structure factor.} The structure factor is defined by \begin{equation}
\begin{array}{c}
S(q,\omega)=\frac{1}{2\pi}\int d\mathbf{r}d\mathbf{r}'dt\left\langle \delta n(\mathbf{r},t)\delta n(\mathbf{r}',0)\right\rangle e^{i\mathbf{q}(\mathbf{r}-\mathbf{r}')-i\omega t}=\\
\\=\frac{N}{2\pi n^{2}}\int dt\left\langle \delta n(\mathbf{q},t)\delta n(-\mathbf{q},0)\right\rangle e^{-i\omega t}\,\,\,,\end{array}\label{41}\end{equation}
where the brackets stand for the thermal average (we leave aside the
central peak). Since \begin{equation}
\delta n(\mathbf{q},t)=-iq\sum_{i}n_{i}u_{i}(\mathbf{q},t)\,\,\,,\label{42}\end{equation}
 it becomes \begin{equation}
S(q,\omega)=\frac{Nq^{2}}{2\pi n^{2}}\int dt\sum_{ij}n_{i}n_{j}\left\langle u_{i}(t)u_{j}(0)\right\rangle e^{-i\omega t}\,\,\,,\label{43}\end{equation}
 where we dropped out the argument $\mathbf{q}$.

In order to calculate the thermal averages we turn back to the system
of equations (\ref{34}) without the external electric field. This
system can be written as \begin{equation}
\begin{array}{c}
(-\omega^{2}+aS_{1})x+bS_{3}y=0\,\,\,,\\
\\aS_{3}x+(-\omega^{2}+bS_{2})y=0\,\,\,,\end{array}\label{44}\end{equation}
where $a=4\pi e^{2}$, $b=\chi q^{2}$, $S_{1,2,3}$ are given by
equation (\ref{35}) and \begin{equation}
x=\frac{1}{n}\sum_{i}z_{i}n_{i}u_{i}\,\,\,,\,\,\, y=\frac{1}{n}\sum_{i}n_{i}u_{i}\,\,.\label{45}\end{equation}
In addition, \begin{equation}
u_{i}=\frac{anz_{i}}{m_{i}\omega^{2}}x+\frac{bn}{m_{i}\omega^{2}}y\,\,.\label{46}\end{equation}
 The system of equations (\ref{44}) has two eigenfrequencies $\omega_{1,2}$
as given by equations (\ref{36}) and (\ref{37}). The corresponding
eigenvectors are given by \begin{equation}
x_{1}\sim S_{1}\,\,,\,\, y_{1}\sim S_{3}\,\,\,;\,\,\, x_{2}\sim bS_{3}\,\,,\,\, y_{2}\sim-aS_{1}\,\,\,\label{47}\end{equation}
in the long wavelength limit. According to equation (\ref{46}) the
coordinates $u_{i}$ can be written as \begin{equation}
u_{i}^{(1,2)}=\frac{anz_{i}}{m_{i}\omega_{1,2}^{2}}x_{1,2}e^{i\omega_{1,2}t}+\frac{bn}{m_{i}\omega_{1,2}^{2}}y_{1,2}e^{i\omega_{1,2}t}\,\,\,,\label{48}\end{equation}
and one can see that they are coordinates of linear harmonic oscillators
with frequencies $\omega_{1,2}$ and potential energies $m_{i}\omega_{1,2}^{2}\left[u_{i}^{(1,2)}\right]^{2}/2$
. The thermal distribution of the coordinate $u$ for such an oscillator
is given by $dw=\sqrt{m\omega^{2}/2\pi T}\exp\left(-m\omega^{2}u^{2}/2T\right)du$
in the classsical limit, where $T$ denotes the temperature ($T\gg\hbar\omega$).
It follows \begin{equation}
\left\langle u_{i}^{(1,2)}u_{j}^{(1,2)}\right\rangle =\frac{T}{m_{i}\omega_{1,2}^{2}}\delta_{ij}\,\,.\label{49}\end{equation}
 Writing \begin{equation}
u_{i}=u_{i}^{(1)}e^{i\omega_{1}t}+u_{i}^{(2)}e^{i\omega_{2}t}\,\,\,\label{50}\end{equation}
 and making use of equation (\ref{49}) the structure factor given
by equation (\ref{43}) becomes \begin{equation}
S(q,\omega)=NTq^{2}\left(\sum_{i}n_{i}^{2}/n^{2}m_{i}\right)\left[\frac{1}{\omega_{1}^{2}}\delta(\omega-\omega_{1})+\frac{1}{\omega_{2}^{2}}\delta(\omega-\omega_{2})\right]\,\,.\label{51}\end{equation}
We can see from this equation that the relevant sound contributions
are given by \begin{equation}
S(q,\omega)\simeq\frac{NT}{v_{s,a}^{2}}\left(\sum_{i}n_{i}^{2}/n^{2}m_{i}\right)\delta(\omega-v_{s,a}q)\,\,.\label{52}\end{equation}

\textbf{Asymmetric short-range interaction.} Up to now, the short-range
interaction was assumed to be the same for all ionic species. In general,
we may introduce a short-range interaction $\chi_{ij}$ depending
on the nature of the ionic species. If this interaction is separable,
the solution given above for a multi-component plasma holds with minor
modifications. For a non-separable short-range interaction, appreciable
changes may appear in the spectrum, which may exhibit multiple branches.
Such a spectrum may serve to identify the nature (mass, charge) of
various molecular aggregates in a multi-component plasma. It is worth
noting that a range of frequencies $10^{10}s^{-1}-10^{12}s^{-1}$
is documented in living cells by microwave, Raman and optical spectroscopies
and by cell-biology studies, upon which the theory of coherence domains
in living matter is built.%
\footnote{See, for instance, H. Frohlich, Phys. Lett. \textbf{A26} 402 (1968);
Int. J. Quant. Chem. \textbf{2} 641 (1968); S. J. Webb, M. E. Stoneham
and H. Frohlich, Phys.Lett \textbf{A63} 407 (1977); S. Webb, Phys.
Reps.\textbf{ 60} 201 (1980); S. Rowlands et al, Phys. Lett. \textbf{A82}
436 (1981); S. C. Roy, Phys. Lett. \textbf{A83} 142 (1981); E. del
Giudice et al, Nucl. Phys. \textbf{B275} 185 (1986).%
} 

We consider here again the $H^{+z}-O^{-2z}$ plasma with different
short-range interaction $\chi_{HH}=\chi_{1\,,\,}\chi_{OO}=\chi_{2}\,,\,\chi_{OH}=\chi_{3}$;
it still exhibits two branches of frequencies, a plasmonic one ($\omega_{1}$)
and a sound-like one ($\omega_{2}$), but the spectrum may have certain
peculiarities (the dielectric constant is not affected by this modification).
Equations of motion (\ref{15}) become now \begin{equation}
\begin{array}{c}
m\ddot{u}+2nq^{2}(\varphi+\chi_{1})u-nq^{2}(2\varphi-\chi_{3})v=-izeq\phi\,\,\,\\
\\M\ddot{v}+nq^{2}(4\varphi+\chi_{2})v-2nq^{2}(2\varphi-\chi_{3})u=2izeq\phi\,\,.\end{array}\label{53}\end{equation}
We introduce the notations \begin{equation}
a=2nq^{2}\varphi/m=8\pi ne^{2}z^{2}/m\,\,\,,\,\,\, b_{1,2,3}=n\chi_{1,2,3}/m\,\,.\label{54}\end{equation}

The dispersion relations can be computed straightforwardly. In the
long wavelength limit ($q\rightarrow0$) we get the plasmonic branch
\begin{equation}
\omega_{1}^{2}\simeq(1+2\alpha)a+\frac{2b_{1}+\alpha^{2}b_{2}-4\alpha b_{3}}{1+2\alpha}q^{2}\,\,\,,\label{55}\end{equation}
where $(1+2\alpha)a=16\pi ne^{2}z^{2}/\mu$ is the plasma frequency,
and the sound-like branch \begin{equation}
\omega_{2}^{2}\simeq\frac{\alpha(4b_{1}+b_{2}+4b_{3})}{1+2\alpha}q^{2}=v_{s}^{2}q^{2}\,\,;\label{56}\end{equation}
 one can see that the sound velocity $v_{s}$ is always a real quantity.
\begin{figure}
\noindent \begin{centering}
\includegraphics[clip,scale=0.4]{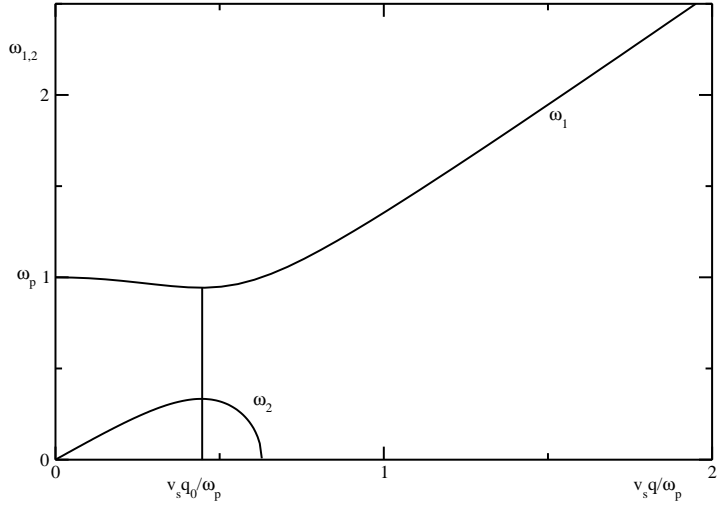}
\par\end{centering}

\caption{Excitation spectrum given by equation (\ref{59}) for the $H^{+z}-O^{-2z}$
plasma with short-range potentials $\chi_{OO}=\chi_{HH}=0$ and $\chi_{OH}=\chi\neq0$.}

\end{figure}

The sound-like branch exhibits an asymptote in the short-wavelength
limit given by \begin{equation}
\omega_{2}^{2}\sim\frac{1}{2}\left[2b_{1}+\alpha b_{2}-\sqrt{(2b_{1}-\alpha b_{2})^{2}+8\alpha b_{3}^{2}}\right]q^{2}\,\,\,,\label{57}\end{equation}
 whose slope may have either sign or vanish. It is easy to see that
this slope is positive for $b_{3}^{2}<b_{1}b_{2}$, negative for $b_{3}^{2}>b_{1}b_{2}$
(when the sound-like branch has a maximum value) and it vanishes for
$b_{3}^{2}=b_{1}b_{2}$ (when the sound-like branch has an horizontal
asymptote). In the case of a negative slope the sound velocity may
exhibit a negative velocity and the sound may suffer a strong absorption
for moderate values of the wavevector, which may indicate an anomalous
or unphysical situation.

We return now to the plasmon branch given by equation (\ref{55}),
and write it as \begin{equation}
\omega_{1}^{2}=\omega_{p}^{2}+b_{2}\frac{2x^{2}-4\alpha\lambda x+\alpha^{2}}{1+2\alpha}q^{2}\,\,\,,\label{58}\end{equation}
 where $\lambda^{2}=b_{3}^{2}/b_{1}b_{2}$ and $x=\sqrt{b_{1}/b_{2}}.$
It is easy to see that for $\lambda^{2}>1$ the plasmonic spectrum
exhibits a dip around a certain value $q_{0}$ of the wavevector $q$
for $\left(\lambda-\sqrt{\lambda^{2}-1/2}\right)\alpha<\sqrt{b_{1}/b_{2}}<\left(\lambda+\sqrt{\lambda^{2}-1/2}\right)\alpha$;
it approaches an asymptote with a positive slope for $q\rightarrow\infty$,
which may define again an anomalous sound for small values of $\omega_{p}$. 

We illustrate these anomalies for a particular case of short-range
interaction $\chi_{1,2}=0$ and $\chi_{3}=\chi$ ($b_{3}=n\chi/m$).
The dispersion relations of the system of equations (\ref{53}) become
\begin{equation}
\omega_{1,2}^{2}=\frac{1}{2}\omega_{p}^{2}\left[1\pm\sqrt{1-4v_{s}^{2}q^{2}/\omega_{p}^{2}+\frac{(1+2\alpha)^{2}}{2\alpha}v_{s}^{4}q^{4}/\omega_{p}^{4}}\right]\,\,.\label{59}\end{equation}
The plasmonic branch has a minimum value for $q_{0}\simeq2\sqrt{m/M}\omega_{p}/v_{s}$,
where the sound-like branch has a maximum value ($\simeq\sqrt{2m/M}\omega_{p}$).
The spectrum is shown in Fig. 2. Using $\omega_{p}\simeq10^{13}s^{-1}$
estimated above and the sound velocity $v_{s}\simeq3000m/s$ in water
we get $q_{0}^{-1}\simeq6\textrm{\AA}$. We may expand $\omega_{1}$
in series of $(q-q_{0})^{2}$ around its mimimum value at $q_{0}$
and get $\omega_{1}\simeq\omega_{p}+(M/4m+1)(v_{s}^{4}q_{0}^{2}/\omega_{p}^{3})(q-q_{0})^{2}=\omega_{p}+(1+4m/M)v_{s}^{2}(q-q_{0})^{2}/\omega_{p}$.
This is similar with the rotons-like dispersion relation discussed
in connection with the coherence domains in water.%
\footnote{G. Preparata, \emph{QED Coherence in Matter}, World Sci (1995). %
} Although this might be an interesting suggestion, it is inconsequential
here, because $\omega_{p}$ is too small in comparison with the temperatures
at which water exists and, therefore, this \char`\"{}dip\char`\"{}
feature has no effect for the water thermodynamics. 

\textbf{Conclusion.} We summarize the main features of the model suggested
here for liquid water. First, we assume, as it is generally accepted,
the four, directional $sp^{3}$-oxygen electronic orbitals. The electron
delocalization along two such orbitals together with a corresponding
delocalization of the hydrogen electronic charge lead to the water
cohesion. It is represented by the cohesion energy $\varepsilon_{0}$
discussed here. Within such a picture, we can still visualize the
oxygen and the hydrogen as neutral atoms, moving around almost freely
(as a consequence of the uniformity of the environment; this gives
a noteworthy support to the \char`\"{}hydrogen bonds\char`\"{} concept).%
\footnote{The point of view taken in this paper is that the hydrogen bonds in
water are introduced in order to account for the uniformity of the
environment of a water molecule in liquid water. As such, it helps
understand the cohesion. However, a consistent upholding of the hydrgen-bonds
concept would mean a vanishing dipole momentum of liqud water. Pauling
himself, (L. Pauling, \emph{loc cit}) who introduced originary this
concept, qualifies it by admiting an asymmetry in the four hydrogen
bonds around an oxygen ion, arising from the two-out-of-four occupied
orbitals. We suggest that the uniformity of the environment makes
the hydrogen atoms (ions) moving as independent entities, while the
asymmetry induces a small charge $z$, so the ion motion is subjected
to Coulomb (and short-range interactions). The electric moment is
ascribed to the directional character of the $sp^{3}$-oxygen electronic
orbitals and the charge transfer between oxygen and hydrogen. Thereby,
the hydrogen-bond concept is employed here through its two features,
directionality and uniformity, with a slight asymmetry, all viewed
as independent qualitative ingredients. %
} To this picture the present model adds another component, arising
from a very small charge transfer between hydrogen and oxygen atoms,
leading to a $H^{+z}-O^{-2z}$ plasma, with the reduced charge $z$.
It may originate in the weak asymmetry of the two occupied $sp^{3}$-oxygen
electronic orbitals with respect to the other two unoccupied orbitals.
Under these circumstances, the hydrogen and oxygen ions interact,
both by Coulomb and short-range potentials. This interaction gives
the plasma frequency and the sound-like excitations frequency. The
plasmons contribute to the excitations which give rise to a consistent
thermodynamics for liquids, in a model introduced recently. In addition,
the ionic plasma oscillations entail oscillations of the delocalized
electronic cloud, with the same eigenfrequency. Subjected to an external
field, these electronic oscillations produce an intrinsic polarizability
which removes the $\omega=0$ singularity in the plasma dielectric
function (the $\omega_{0}$ frequency). In addition, the magnitude
of the electric moment $\mathbf{p}$ which is responsible for the
orientational, static dielectric function is in satisfactory agrement
with the plasma charge $z$ derived herein. 

On the basis of this model we are able to understand to some extent,
both qualitatively and in some places even quantitatively, the sound
anomaly, the dielectric function (permitivity dispersion), the structure
factor, cohesion and thermodynamics of water. The model is extended
to a multi-component classical plasma, including an asymmetric short-range
interaction between the components, which might be relevant for more
complex structural aggregates like those in biological matter. 
\end{document}